\documentclass[11pt,a4paper]{article} 
\pdfoutput=1

\usepackage[noend]{algpseudocode}
\usepackage{algorithm}
\usepackage{amsmath,amsfonts,amsbsy,amssymb,array,accents}
\usepackage{enumerate,array,latexsym,graphicx,mathrsfs,verbatim,psfrag}
\usepackage{bm} 
\usepackage[normalem]{ulem}
\usepackage{booktabs} 
\usepackage[usenames]{color}

\usepackage[english]{babel}
\usepackage{fancybox,url}

\usepackage[nosort]{cite}
\usepackage{chngpage} 

\usepackage[utf8x]{inputenc}

\usepackage{enumitem}
\setlist{itemsep=0pt}

\usepackage{algorithm}
\usepackage{algpseudocode}

\usepackage[colorlinks=true,      linkcolor=darkblue,      urlcolor=darkblue,      
            filecolor=darkblue,      citecolor=darkblue,       pdfstartview=FitH,     
						pdfpagemode=UseNone,      bookmarksopen=true]{hyperref}  
\usepackage[all]{hypcap}     

\usepackage{mathtools}

\DeclarePairedDelimiterX\braket[2]{\langle}{\rangle}{#1 \delimsize\vert #2}

\newcommand{\captionfonts}{\small}
\makeatletter  
\long\def\@makecaption#1#2{%
  \vskip\abovecaptionskip
  \sbox\@tempboxa{{\captionfonts #1: #2}}%
 \ifdim \wd\@tempboxa >\hsize
    {\captionfonts #1: #2\par}
  \else
    \hbox to\hsize{\hfil\box\@tempboxa\hfil}%
  \fi
  \vskip\belowcaptionskip}
\makeatother   

\DeclareMathSymbol{\medhatsym}{\mathord}{largesymbols}{"62} 

\DeclareMathSymbol{\medtildesym}{\mathord}{largesymbols}{"65}

\mathchardef\mhyphen="2D

\def\({\left(}
\def\){\right)}
\def\[{\left[}
\def\]{\right]}

\def\barray{\begin{array}}
\def\earray{\end{array}}
\def\be{\begin{equation}}
\def\ee{\end{equation}}
\def\bea{\begin{eqnarray}}
\def\eea{\end{eqnarray}}
\def\bal{\begin{align}}
\def\eal{\end{align}}

\def\ba{\begin{aligned}}
\def\ea{\end{aligned}}

\numberwithin{equation}{section} %

\makeatletter
\g@addto@macro\bfseries{\boldmath}
\makeatother

\definecolor{cardinal}{rgb}{0.6,0,0}
\definecolor{darkgreen}{rgb}{0,0.4,0}
\definecolor{purple}{rgb}{0.5, 0, 0.5}
\definecolor{golden}{rgb}{0.92, 0.7, 0}
\definecolor{midnight}{rgb}{0, 0, 0.5}
\definecolor{darkblue}{rgb}{0, 0, 0.8}

\def\cM{{\cal M}}

\begin{document}

\begin{flushright}
%
%
\end{flushright}

\vspace{19mm}

\begin{center}

{\huge \bf{Evolutionary algorithms for\\[3mm] multi-center solutions}}

\vspace{22mm}

{\large
\textsc{Sami Rawash, ~David Turton}
}

\vspace{15mm}

Mathematical Sciences and STAG Research Centre,\\ University of Southampton,\\
Highfield, Southampton SO17 1BJ, UK

\vspace{7mm}

{\footnotesize\upshape\ttfamily  s.rawash @ soton.ac.uk, ~d.j.turton @ soton.ac.uk } \\

\vspace{20mm}

\textsc{Abstract}
\vspace{7mm}
\begin{adjustwidth}{12mm}{12mm} 
\noindent
Large classes of multi-center supergravity solutions have been constructed in the study of supersymmetric black holes and their microstates. 
Many smooth multi-center solutions have the same charges as supersymmetric black holes, with all centers deep inside a long black-hole-like throat.
These configurations are constrained by regularity, absence of closed timelike curves, and charge quantization.
Due to these constraints, constructing explicit solutions with several centers in generic arrangements, and with all parameters in physically relevant ranges, is a hard task. 
In this work we present an optimization algorithm, based on evolutionary algorithms and Bayesian optimization, that systematically constructs numerical solutions satisfying all constraints.
We exhibit explicit examples of novel five-center and seven-center machine-precision solutions.
\end{adjustwidth}

\end{center}

\thispagestyle{empty}

\newpage


%
%


\baselineskip=15pt
\parskip=3pt

\setcounter{tocdepth}{2}
\tableofcontents
\section{Introduction}

Black hole solutions in classical gravitational theories typically involve a small handful of parameters, such as mass, angular momentum, and charge. However black holes have an entropy proportional to their horizon area, which suggests that they have a vast number of internal degrees of freedom. Black holes also contain curvature singularities, and the semiclassical description of black hole evaporation leads to the information paradox~\cite{Hawking:1976ra,Mathur:2009hf}. These facts present three corresponding major challenges for a fundamental theory of quantum gravity: to identify the black hole internal degrees of freedom, to resolve the singularities inside black holes, and to provide a consistent description of black hole evaporation.

In String Theory, black hole entropy arises from an exponential number of internal quantum microstates~\cite{Strominger:1996sh}. It is therefore of significant interest to study the gravitational description of heavy pure states, in order to investigate string-theoretic singularity resolution and black hole evaporation. Large families of such pure states are well-described by smooth, horizonless supergravity solutions which, in the best-understood examples, provide a valuable description of black hole microstates~\cite{Lunin:2001jy,Mathur:2005zp,Skenderis:2008qn,Balasubramanian:2008da}.

In recent years, an increasing number of String Theory, and Particle Physics, problems have been addressed with optimization algorithms and machine learning, see e.g.~\cite{Abel:2022nje,Shanahan:2022ifi,Bena:2021wyr, Ruehle:2020jrk, Abel:2014xta, Partipilo:2022alb,Erbin:2022rgx,Berglund:2022gvm}.
In this work we present an algorithm, based on evolutionary algorithms and Bayesian optimization, to construct smooth horizonless supergravity solutions.
The category of microstate solutions that we study are
supersymmetric multi-center solutions, also known as bubbling solutions~\cite{Bena:2004de,Gauntlett:2004qy,Bena:2006kb,Bena:2007kg,Bena:2007qc,Avila:2017pwi,Heidmann:2017cxt, Bena:2017fvm, Bianchi:2017bxl, Mayerson:2022yoc}.

Supersymmetric multi-center solutions involve non-trivial topology supported by flux. 
These solutions are specified by a set of harmonic functions on a three-dimensional Euclidean space.
This formalism is typically used to construct supergravity fields in four, five or six macroscopic dimensions. 
Depending on the details, large families of multi-center solutions in five and/or six macroscopic dimensions can have the features of being horizonless and smooth, up to possible orbifold singularities, see e.g.~\cite{Bena:2007kg}.
We shall moreover focus on the ``scaling'' regime in which the centers lie deep inside a long black-hole-like throat~\cite{Bena:2006kb,Bena:2007kg,Bena:2007qc}.

Multi-center solutions are one of two main classes of smooth horizonless supersymmetric solutions. The other class is known as {\it superstrata}~\cite{Bena:2015bea,Bena:2016agb,Bena:2016ypk,Bena:2017geu,Tyukov:2017uig,Bena:2017xbt,Ceplak:2018pws,Heidmann:2019zws,Heidmann:2019xrd,Ganchev:2022exf,Ceplak:2022pep}. Of the two classes, superstrata have a proposed holographic description that has passed precision holographic tests, involving protected correlators that can be reliably compared across moduli space between supergravity and the dual symmetric product orbifold CFT~\cite{Kanitscheider:2006zf,Kanitscheider:2007wq,Giusto:2015dfa,Giusto:2019qig,Rawash:2021pik, Ganchev:2021ewa}. Two-center bubbling solutions have a similarly well-established holographic description~\cite{Giusto:2004id,Giusto:2004ip,Lunin:2004uu,Jejjala:2005yu,Giusto:2012yz,Chakrabarty:2015foa} and also a string worldsheet description~\cite{Martinec:2017ztd,Martinec:2018nco,Martinec:2019wzw,Martinec:2020gkv,Bufalini:2021ndn,Bufalini:2022wyp,Bufalini:2022wzu,Martinec:2022okx}. By contrast, multi-center solutions with three or more centers do not have a proposed holographic description, and it has been argued that they do not describe microstates of a single supersymmetric black hole, though they could describe microstates of other black objects~\cite{Bossard:2019ajg}. 
Nevertheless, multi-center solutions provide interesting examples of gravitational solutions that closely resemble black holes, especially in the scaling regime. Indeed, multi-center solutions have been used to investigate potentially observable signatures of string theoretic black hole microstructure in gravitational wave observations~\cite{Bena:2020see,Bianchi:2020bxa,Bah:2021jno}.

Constructing multi-center solutions with several centers is a hard problem. This is because asymptotic flatness, charge quantization, smoothness and absence of closed timelike curves (CTCs) comprise a set of non-trivial algebraic constraints. These constraints make the positions of the centers and the coefficients of the poles of the harmonic functions highly interdependent. The most important set of these constraints is known as the bubble equations. 
The distances between the centers are generically irrational real quantities. 
The bubble equations constrain these distances in terms of quantized parameters; this is a strong set of constraints. For further discussion, see e.g.~\cite{Avila:2017pwi}.

Despite this difficulty, several solutions with three or four centers have been constructed analytically, see e.g.~\cite{Bena:2006kb,Bena:2007qc,Heidmann:2017cxt, Bena:2017fvm, Bianchi:2017bxl, Mayerson:2022yoc}. However, fewer solutions with five or more centers have been constructed, and until relatively recently these typically involved taking all centers to lie on a line, so that the configuration is axisymmetric, see e.g.~\cite{Bena:2006kb}.
An important step forward was recently made, by considering the dependent variables in the bubble equations to be a subset of the coefficients of the poles of the harmonic functions (which we call ``flux parameters''), rather than the distances between the centers~\cite{Avila:2017pwi}. In this form, the bubble equations are a linear system, involving a symmetric matrix $\mathcal{M}$.
It was moreover conjectured that a configuration does not contain CTCs if and only if $\mathcal{M}$ is positive-definite.
While this perspective simplifies the task of finding physically relevant solutions, the strong nature of the constraints between generically irrational distances and quantized parameters remains.

One can construct exact solutions with this method by arranging non-generic locations of centers, however this is not easy to implement in practice, and is currently limited to quite special arrangements of centers~\cite{Avila:2017pwi}. An alternative approach is to construct approximate solutions to the bubble equations, as discussed in~\cite{Avila:2017pwi} and done in~\cite{Heidmann:2017cxt,Bena:2017fvm}. One can do so in an iterative approach by first choosing a set of locations of the centers and  solving for the flux parameters, obtaining generically irrational values.  One then rounds any irrational flux parameters to nearby rational values that enable all quantization constraints to be satisfied. One then takes the rounded fluxes and attempts to re-solve the bubble equations in the traditional approach, to find the distances between the centers. This method has been used to numerically construct four-center solutions in axisymmetric or near-axisymmetric configurations~\cite{Bena:2017fvm}.
However, it is unclear whether this method is generically tractable for more than four centers~\cite{Avila:2017pwi}.

In this paper we present a novel algorithmic method to construct numerical solutions with any number of centers, and with no symmetry imposed on the locations of the centers.
The basic idea is to proceed in two steps. First, we generate a suitably good starting configuration which has certain desired physical properties, but is not yet a solution to the bubble equations. Second, we systematically vary the positions of the centers to construct a sequence of approximate solutions with increasing precision.

These two steps require two separate algorithms, since they optimize over different variables. 
In the first step, we generate a good starting configuration by optimizing over flux parameters. In the second step, we optimize over the locations of the centers.

The starting configuration is required to have two key physical properties. The first requirement is that the configuration be in the scaling regime mentioned above, in which the centers lie deep inside a long black-hole-like throat. We implement this by first finding an exact solution to the {\it homogeneous} form of the bubble equations, which will be reviewed in Section~\ref{sec: problem}. We then round the flux parameters as described above.

The second requirement is that the supergravity charge radii are large, such that the solutions are weakly curved. We shall primarily have in mind solutions corresponding to bound states of D1 branes, D5 branes, and momentum P in a compact direction, in five or six macroscopic dimensions. In six dimensions,
the (dimensionful) D1 and D5 charges,  $Q_1,Q_5$ control the main curvature scale of the solution (and contribute to the ADM mass), so we require them to be appropriately large. 

The charges $Q_1,Q_5$, considered as functions of
the flux parameters, are computationally expensive to evaluate. Bayesian optimization is well-suited to the task of optimizing computationally expensive functions (see e.g.~\cite{Frazier2018ATO}); we therefore employ it for the first step.
The reason that the charges are computationally expensive to evaluate is that, given some flux parameters, one must first solve the bubble equations for the remaining flux parameters, and then evaluate the expressions for the charges $Q_1$, $Q_5$, which will be given in Section~\ref{sec: problem}.

In a nutshell, Bayesian optimization is a strategy to choose points (in our case, values of flux parameters) on which to evaluate, or sample, the function to be optimized, known as the ``objective'' function (in our case, $\min(Q_1,Q_5)$). 
After a point is sampled, the accrued knowledge of the objective function is updated, and then used to decide the next point to sample. The next point is selected according to a specified strategy that balances exploitation of more favourable regions (where $Q_1,Q_5$ are known to be larger) versus exploration of unknown regions (where $Q_1,Q_5$ have not yet been computed). We will describe this in detail in Section~\ref{sec: BO}.

After a successful run of the Bayesian optimization algorithm, we have a configuration in the scaling regime, with appropriately large charges $Q_1$, $Q_5$, which is not yet a solution. It can be regarded as an approximate solution, but with low precision. In the second step, we construct numerical solutions by varying the positions of the centers.

Our two-step approach means that after a successful first step, it is reasonable to expect that if there is a genuine solution nearby, it is likely to require incremental adjustments to the positions of the centers, rather than a wide search. 
Evolutionary algorithms are well-suited to problems where incremental changes result in incremental improvements (see e.g.~\cite{Abel:2014xta}). We therefore employ an evolutionary algorithm in the second step.

Evolutionary algorithms work by generating a population of {\it individuals}, comprising certain data known as {\it genes}, and quantifying their  {\it fitness} via a function known as the fitness function. 
The algorithm then generates subsequent generations of individuals  following the principles of the Darwinian theory: selection, reproduction and mutation. By iterating this process over several generations, the algorithm aims to construct new individuals with higher fitness.

In our algorithm, an individual is a multi-center supergravity configuration that approximately solves the bubble equations (in their full, {\it inhomogeneous} form).  Its genes are (an appropriate subset of) the positions of the centers.  The coefficients of the poles of the harmonic functions are determined by the previous step. The fitness function quantifies the precision to which the bubble equations are approximately satisfied, with higher fitness being an approximate solution with a lower error.
Once a multi-center configuration with the desired fitness is generated, we investigate the absence of CTCs by computing the eigenvalues of the matrix $\mathcal{M}$.

Our algorithm is designed to generate solutions with any number of centers, in a generic configuration. 
We have run the algorithm on several configurations of three, five and seven centers, and we shall exhibit explicit examples of novel five- and seven-center configurations. Generating solutions with a higher number of centers is feasible, although naturally this is computationally more expensive. The algorithm is implemented in Python, and the code is publicly available.\footnote{GitHub URL: \url{https://github.com/SamiRawash/Multicenter-Scaling-Solutions}.}

This paper is structured as follows. In Section~\ref{sec: problem} we review multi-center scaling solutions, and their construction in the formalism of~\cite{Avila:2017pwi}. In Section~\ref{sec:algorithm} we first describe our overall method, and then describe the Bayesian optimization algorithm and the evolutionary algorithm we have developed. In Section~\ref{results} we describe explicit examples of five-center and seven-center scaling configurations obtained with our method, and comment on the performance of the algorithm. We discuss our results in Section~\ref{sec: conclusions}.

\section{Multi-center scaling supergravity solutions}\label{sec: problem}

\subsection{Multi-center solutions}

For concreteness, we primarily consider 5D $\mathcal{N}=1$ Super-Einstein-Maxwell-Yang-Mills supergravity, whose bosonic field content is the metric, three Abelian vector multiplets, and an SU(2) triplet of non-Abelian vector multiplets. If one turns off the non-Abelian multiplets, one recovers the STU model.

Multi-center solutions are specified by a set of harmonic functions on a three-dimensional Euclidean ``base'' space, which have poles at the location of the centers. The index $a=0,1,...,n-1$  labels the centers, and $r_a=|\vec r - {\vec r}_a|$ is the distance from the $a$-th center in the three-dimensional base. 
In the Abelian sector the harmonic functions  are ($i=0,1,2$):
\begin{equation}\label{Harmonic functions 1}
    H\,=\,\sum_{a=0}^{n-1} \frac{q_a}{r_a}\,,\qquad
    K^i\,=\,\sum_{a=0}^{n-1} \frac{k^i_a}{r_a}\,,\qquad
    L^i\,=\,l_0^i + \sum_{a=0}^{n-1} \frac{l^i_a}{r_a}\,,\qquad
    M\,=\, m_0 +\sum_{a=0}^{n-1} \frac{m_a}{r_a}\,,
\end{equation}
where $q_a \in \mathbb{Z}$. In the non-Abelian sector~\cite{Ramirez:2016tqc},
denoting the gauge coupling by $g$, we have
\begin{equation}\label{Harmonic functions 2}
    P\,=\,1+\sum_{a=0}^{n-1} \frac{\lambda_a}{r_a}\,,\qquad
    Q\,=\,\sum_{a=0}^{n-1} \frac{\sigma_a\lambda_a}{r_a}\,.
\end{equation}
The harmonic function $H$ defines a four-dimensional Gibbons-Hawking metric via
\begin{equation}\label{ds4 GH}
ds_4^2 \,=\,H^{-1}(d\psi+A)^2+Hds^2_3   \;, 
\end{equation}
where $ds^2_3$ is the flat metric on $\mathbb{R}^3$, and $A$ is a one-form related to $H$ via $\;\star_3\,dA=dH$. For the full five-dimensional fields, we refer the reader to~\cite{Avila:2017pwi}.

Only certain subsets of possible coefficients of the poles in Eqs.~\eqref{Harmonic functions 1} and~\eqref{Harmonic functions 2} lead to physically sensible solutions: one needs to impose further constraints. First, asymptotic flatness requires  $\sum_a q_a=1$. Second, upon uplifting to Type IIB supergravity compactified on $S^1 \times T^4$, the coefficients $k^i_a$ are quantized in terms of integer flux parameters $n^i_a$ as follows~\cite{Giusto:2012yz},
\begin{equation}\label{k quantization condition}
    k^0_a\,=\,\frac{g_s\alpha'}{2R_y}n^0_a\,,\qquad
    k^1_a\,=\,\frac{g_s\alpha'^3}{2V_{4}R_y}n^1_a\,,\qquad
    k^2_a\,=\,\frac{R_y}{2}n^2_a\,,
\end{equation}
where the coordinate volume of $T^4$ is $(2\pi)^4V_{4}$ and that of the $S^1$ is $2\pi R_y$. 

In this paper we focus on smooth horizonless supersymmetric solutions.\footnote{The supersymmetric multi-center formalism can also be used to construct solutions with physical singularities such as shockwaves~\cite{Chakrabarty:2021sff}, which give collective descriptions of families of pure states. Similar but different multi-center formalisms exist for non-supersymmetric solutions~\cite{Bena:2016dbw,Bossard:2017vii,Bah:2021owp,Bah:2022yji}.} The following relations are imposed by absence of event horizons and singularities (the first three relations) and asymptotic flatness (the last two relations), see e.g.~\cite[App.~A.3]{Avila:2017pwi},
\begin{equation}
\begin{aligned}
l^i_a\, &= \, -\frac{|\epsilon^{ijk}|}{2}\frac{k_a^j k_a^k}{q_a}+\frac{\delta^{0i}}{2g^2}\,,\qquad
\sigma_a\, = \, \frac{k^0_a}{q_a}\,, \qquad m_a\, = \, \frac{k_a^0}{2q_a^2}\big(k^1_ak^2_a - \frac{1}{2g^2}\big)\,,\\
l^0_0l^1_0l^2_0\, &= \, 1\,, \qquad m_0\, = \, -\frac{1}{2}\sum_{a,i}l^i_0k^i_a\,.
\end{aligned}
\end{equation}

The absence of Dirac-Misner singularities imposes the so-called ``bubble equations''~\cite{Denef:2000nb,Bena:2007kg,Ramirez:2016tqc}, which constrain the relation between the positions of the centers and the local charges:
\begin{equation}\label{bubble equation original}
    \sum_{b\neq a}\frac{q_a q_b}{r_{ab}}\Pi^0_{ab}\Big(\Pi^1_{ab}\Pi^2_{ab}-\frac{1}{2g^2} \mathbb{T}_{ab}\Big)=\sum_{b,i} q_aq_bl^i_0\Pi^i_{ab}\,,
\end{equation}
where 
\begin{equation}
    \Pi^i_{ab}=\frac{k^i_b}{q_b}-\frac{k^i_a}{q_a}, \qquad \mathbb{T}_{ab}=\frac{1}{q_a^2}+\frac{1}{q_b^2}\,.
\end{equation}
Here $r_{ab}$ is the $\mathbb{R}^3$ Euclidean distance between centers $a$ and $b$, $\Pi^i_{ab}$ are the magnetic fluxes, and we will refer to the coefficients $k^i_a$ as flux parameters. The bubble equations are a set of $n$ equations among which only $(n-1)$ are independent: summation over $a$ leads to a trivial identity, due to the antisymmetry of the $\Pi^i_{ab}$.

The asymptotic charges of the multi-center solutions are~\cite{Bena:2007kg}:\footnote{We use conventions in which $J_L$ and $J_R$ are interchanged with respect to~\cite{Bena:2007kg}.}
\begin{equation}\label{global charges}
    \begin{aligned}
        Q_1\,&=\, -\sum_{a,b,c} q_aq_bq_c \Pi^1_{ab}\Pi^2_{ac}+
        \frac{1}{2g^2}\sum_a\frac{1}{q_a}\,,\\
        Q_5\,&=\, -\sum_{a,b,c} q_aq_bq_c \Pi^0_{ab}\Pi^2_{ac}\,,\\
        Q_p\,&=\, -\sum_{a,b,c} q_aq_bq_c \Pi^0_{ab}\Pi^1_{ac}\,,\\
        J_L\,&=\, -\frac12 \sum_{a,b,c,d} q_a q_b q_c q_d \Pi^0_{ab}\Pi^1_{ac}\Pi^2_{ad} +
        \frac{1}{4g^2} \sum_{a,b} \frac{q_b \Pi^0_{ab}}{q_a}\,,\\
        \vec{J}_R\,&\,=\frac{1}{4}\sum_{a,b,a\neq b} q_a q_b \Pi^0_{ab}\Big(\Pi^1_{ab}\Pi^2_{ab}-\frac{1}{2g^2}\mathbb{T}_{ab}\Big)\frac{\vec{r}_a-\vec{r}_b}{|\vec{r}_a-\vec{r}_b|}\,.
    \end{aligned}
\end{equation}

\subsection{Scaling solutions and their construction}

Of particular interest are solutions to the bubble equations~\eqref{bubble equation original} in which the distances between the centers can be made uniformly parametrically small by scaling $r_{ab}\rightarrow \lambda r_{ab}$ with $\lambda \ll 1$, while keeping the asymptotic charges approximately constant. These solutions are known as ``scaling'' solutions~\cite{Bena:2006kb,Bena:2007kg,Bena:2007qc}. 
Note that 
the rescaling $r_{ab}\rightarrow \lambda r_{ab}$ is equivalent to multiplying the RHS of~\eqref{bubble equation original} by $\lambda$, with $\lambda \ll 1$.  
It will be useful for us to note that in the limit $\lambda\to 0$, one obtains the homogeneous bubble equations~\cite{Bena:2007qc} (see also for instance~\cite{Bena:2017fvm}),
\begin{equation}\label{bubble equation AdS}
    \sum_{b\neq a}\frac{q_a q_b}{r_{ab}}\Pi^0_{ab}\left(\Pi^1_{ab}\Pi^2_{ab}-\frac{1}{2g^2} \mathbb{T}_{ab}\right)\,=\, 0\,.
\end{equation}
Therefore, in the scaling regime of small $\lambda$, solutions to the full inhomogeneous bubble equations~\eqref{bubble equation original} are also approximate solutions to the homogeneous bubble equations~\eqref{bubble equation AdS}, up to terms of order $\lambda$. We will exploit this to construct new scaling solutions.

The full inhomogeneous bubble equations~\eqref{bubble equation original} have typically been considered as equations in which the variables to be solved for are the distances $r_{ab}$, see e.g.~\cite{Bena:2007kg}. This perspective has two disadvantages~\cite{Avila:2017pwi}. First, it is generically difficult to find solutions for $r_{ab}$. Second, after solving the equations, one often finds that the resulting $r_{ab}$ do not represent possible distances between points in 3D Euclidean  space; for instance, the triangle inequality might not be respected.

A recently developed alternative approach is to exploit the feature that the bubble equations~\eqref{bubble equation original} are linear in the flux parameters $k^2_a$. Thus, instead of solving for the distances, one can first specify the positions of the centers, and then solve for the flux parameters $k^2_a$ with $a=2,3,...n$~\cite{Avila:2017pwi}\footnote{The bubble equations are also linear in $k^{0,1}_a$, so a similar analysis can be carried out for them.}.
While this procedure is general and not restricted to scaling solutions, let us now review it in the context of scaling solutions. We introduce a scaling parameter $\lambda$ that rescales the positions of the centers while keeping the shape of the distribution fixed: i.e.~we write the distance between the centers as $r_{ab}=\lambda d_{ab}$, where $d_{ab}$ remain constant in the scaling process. We define\footnote{A numerical typo in~\cite[Eq.~(3.22)]{Avila:2017pwi} has been corrected.}
\begin{equation}
    \begin{aligned}
    \label{eq:a-b}
        \bar A_{ab}^2\,&=\, \frac{q_aq_b}{d_{ab}}\Pi^0_{ab}\Pi^1_{ab}\;,\qquad \dot A_{ab}^2\,=\,- s q_a q_b l^2_0\,,\\
        \bar B^2_{ab}\,&=\, \sum_{b=0}^{n-1}\frac{q_a q_b}{d_{ab}}\frac{1}{2g^2}\mathbb{T}_{ab}\Pi^0_{ab}\;,\qquad 
        \dot B^2_{a}\,=\,s \sum_{b=0}^{n-1}q_a q_b (l^0_0\Pi^0_{ab}+l^1_0\Pi^1_{ab})\,,
    \end{aligned}
\end{equation}
where we have introduced the constant $s$ which takes values $0$ or $1$. These values correspond respectively to the homogeneous and inhomogeneous bubble equations, as we shall see momentarily. 
We then introduce ($\alpha\,, \beta\,=\, 1,...,n-1$)\footnote{To be clear, the `2' are superscript labels for the value of the index $i$, not exponents. To avoid potential confusion on this point, we have suppressed the superscript `2' on the matrix $\cM^2$ in the Introduction and Discussion sections.}
\begin{equation}
    \begin{aligned}
        \bar{\mathcal{M}}^2_{\alpha \beta}\,&=\, \bar A^2_{(\alpha+1)(\beta+1)}-\delta^{\beta}_{\alpha}\sum_{c=0}^{n-1}\bar A^2_{(\alpha+1)c}\,,\\
         \dot{\mathcal{M}}^2_{\alpha \beta}\,&=\, \dot A^2_{(\alpha+1)(\beta+1)}-\delta^{\beta}_{\alpha}\sum_{c=0}^{n-1}\dot  A^2_{(\alpha+1)c}\,,
    \end{aligned}
\end{equation}
in terms of which, we write the following linear system of equations in the fluxes $\Pi^2_{ab}$:
\begin{equation}\label{Bubble eq linear}
    \begin{aligned}
        \mathcal{M}^2_{\alpha \beta}\Pi^2_{1(\alpha+1)}\,\equiv\, \Big(\bar{\mathcal{M}}^2_{\alpha \beta}+\lambda\dot{\mathcal{M}}^2_{\alpha\beta}\Big)\Pi^2_{1(\alpha+1)}\, = \, \bar B^2_{\beta}+\lambda \dot B^2_{\beta}\,.
    \end{aligned}
\end{equation}
For $s=1$ this linear system is equivalent to the inhomogeneous bubble equations~\eqref{bubble equation original}, while for $s=0$ the system is equivalent to the homogeneous bubble equations~\eqref{bubble equation AdS}.

Although this perspective has simplified the task of solving the bubble equations, it remains a fact that generic solutions obtained in this way will not respect the quantization conditions in Eq.~\eqref{k quantization condition}. This can be seen as follows. If we choose generic locations of the centers, generic relative distances will be irrational numbers. Then generic solutions will give irrational values of the flux parameters $k^2_\alpha$, which is in conflict with the quantization conditions in Eq.~\eqref{k quantization condition}.

As described in the Introduction, using this method one can construct exact solutions with quantized fluxes by arranging a set of non-generic locations of centers, such that all relative distances are rational. For instance one can take all centers to lie on a line, or on a circle, as discussed in~\cite{Avila:2017pwi}.\footnote{For earlier examples of solutions with all centers on a line, see e.g.~\cite{Bena:2006kb}.} While these constructions provide interesting and valuable exact solutions, the requirement to work with non-generic locations of centers is a significant limitation.

To proceed further, an alternative approach is to construct approximate solutions to the bubble equations. One can do so with an iterative approach, as follows. One first chooses a set of center locations, then solves for $k^2_\alpha$, generically obtaining irrational values. One then rounds the $k^2_\alpha$ to nearby rational numbers to a desired precision, obtaining an approximate solution, as discussed in~\cite{Avila:2017pwi} and done in~\cite{Heidmann:2017cxt,Bena:2017fvm}. 

This can be further improved by taking the rounded flux parameters $k^2_\alpha$, re-solving the bubble equations (in the traditional way) to obtain a new set of distances $r_{ab}$, and then arranging center positions to have the resulting relative distances. 
If this could be done analytically, one can obtain exact solutions, however typically this is hard, for the reasons discussed below Eq.~\eqref{bubble equation AdS}. More realistically, one can employ this method to improve the precision of the approximate solution, as done in~\cite{Bena:2017fvm}.
However, for more than four centers, doubt has been expressed as to the feasibility of this method~\cite{Avila:2017pwi}.

\section{Constructing numerical scaling solutions}\label{sec:algorithm}

\subsection{Overview of the method}
\label{sec:algo-overview}

In this section we describe our method to construct numerical solutions. 
The method involves two main steps. In the first step, we generate an approximate scaling solution with rounded flux parameters $k^2_\alpha$ and large charges $Q_1, Q_5$. The condition of large charges is imposed using a Bayesian optimization algorithm, optimizing over flux parameters. In the second step, we fix the flux parameters and optimize instead over the locations of the centers, using an evolutionary algorithm.

In this subsection we first describe the overall method, focusing primarily on the first step of generating an appropriately good starting configuration. The following subsections will describe the respective algorithms in detail.

A key ingredient in the first step is the construction of configurations in the scaling regime. To do this, we follow the discussion below~\eqref{bubble equation AdS}. We solve the homogeneous bubble equations~\eqref{bubble equation AdS} in the form~\eqref{Bubble eq linear} by taking the position of the centers $\vec r_a$ and the coefficients $q_a$, $l_0^i$, $k^{0,1}_a$ and $k^{2}_{0}$ to be independent variables, and solving for $k^{2}_{\alpha}$ (recall $\alpha= 1,...,n-1$).
The initialization/optimization of the independent variables will be described in the next subsection.

This will enable us, at a later step, to rescale $r_{ab}\rightarrow \lambda r_{ab}$ with $\lambda \ll 1$ to obtain a configuration in the scaling regime, as done in~\cite{Bena:2017fvm}.\footnote{We thank Pierre Heidmann for a discussion on this point.} However, before doing this rescaling we round the flux parameters, and impose $Q_1, Q_5 > \bar Q$, as follows.

We round the flux parameters, $k^2_{\alpha}\rightarrow \tilde k^2_{\alpha}$, to a certain precision, which will be a hyperparameter of the algorithm, and which we call \texttt{k\_rounding}. From here onwards, tildes denote rounded quantities. After rounding, Eq.~\eqref{bubble equation AdS} will no longer be exactly satisfied for the same configuration of centers.

Having constructed an approximate solution to the homogeneous bubble equations, we then impose $Q_1, Q_5 > \bar Q$ using a Bayesian optimization algorithm, optimizing over a subset of the independent flux parameters, and iterating over the steps so far, as described in the next subsection.

After a successful run of the Bayesian optimization algorithm, we 
have an approximate solution to the homogeneous bubble equations, with charges $Q_1,Q_5$ in the desired range. We next generate another approximate solution in the scaling regime by rescaling the positions of the centers obtained in Eq.~\eqref{r uniform distribution}:
\begin{equation}\label{r rescaled}
    r^i_a\rightarrow\,\,\bar r^i_a\,\equiv\,\lambda\,r^i_a\,,\qquad \lambda \ll 1 \,,
\end{equation}
where for concreteness we take $\lambda\,=\,10^{-5}$.

At this stage, we have a starting configuration with the desired physical properties. Before using it as an input to the evolutionary algorithm, we next impose two conditions that further indicate whether the configurations can be considered sufficiently good starting configurations.

To describe the first condition, let us consider the homogeneous bubble equations~\eqref{bubble equation AdS} for $a=n-1\,$:
\begin{equation}\label{bubble equation AdS2}
    \sum_{b\neq n-1}\frac{q_{n-1} q_b}{r_{(n-1)b}}\Pi^0_{(n-1)b}\left(\Pi^1_{(n-1)b}\Pi^2_{(n-1)b}-\frac{1}{2g^2} \mathbb{T}_{(n-1)b}\right)\,=\, 0\,.
\end{equation}
Let us further examine the generic case in which all the terms in this sum are non-zero. Then a necessary condition to have a solution is that not all of the terms in the sum have the same sign. Since the distances are positive, the expressions  $q_{n-1} q_b \Pi^0_{(n-1)b}(\Pi^1_{(n-1)b} \Pi^2_{(n-1)b}-\frac{1}{2g^2} \mathbb{T}_{(n-1)b})$ 
should not all have the same sign.

After rounding the flux parameters, $k^2_{\alpha}\rightarrow \tilde k^2_{\alpha}$, we have rounded fluxes $\tilde\Pi^2$. In the evolutionary algorithm, we will keep these fixed and change the location of the centers. So, before running the evolutionary algorithm, we examine whether or not all of the following expressions have the same sign (note the presence of the rounded fluxes $\tilde\Pi^2$):
\begin{equation}\label{flux signs}
q_{n-1} q_b \Pi^0_{(n-1)b}\left(\Pi^1_{(n-1)b} \tilde\Pi^2_{(n-1)b}-\frac{1}{2g^2} \mathbb{T}_{(n-1)b}\right) .
\end{equation}
We have found that if these quantities have the same sign, it is a reliable indicator that there is unlikely to be a nearby scaling solution. Therefore, only if these quanties do not all have the same sign, we proceed.

Next, we perform a preliminary investigation of the absence of CTCs.
Let us first review the case in which only Abelian fields are turned on. To rule out CTCs,
two algebraic combinations of the harmonic functions~\eqref{Harmonic functions 1} must be globally positive.
Generically, the stronger of these conditions is that the quartic $E_{7(7)}$ invariant, as a function of the harmonic functions~\eqref{Harmonic functions 1}, is globally positive~\cite{Bena:2005va}. Investigating this condition is non-trivial, and typically done numerically.
The generalization to configurations with both Abelian and non-Abelian fields was discussed in~\cite{Ramirez:2016tqc,Avila:2017pwi}. The authors of~\cite{Avila:2017pwi} conjectured that the condition for absence of CTCs is equivalent to requiring that the matrix $\mathcal{M}^2$ defined in Eq.~\eqref{Bubble eq linear} is positive-definite. We thus investigate absence of CTCs in the solutions found by the algorithm by examining this condition on $\mathcal{M}^2$.

Although $\mathcal{M}^2$ depends on the positions of the centers, and thus will be modified by the evolutionary algorithm, we have observed that small modifications of the distances do not tend to change the eigenvalues much. Of course, after the evolutionary algorithm, one must recheck the condition on $\mathcal{M}^2$. However, checking the condition at this stage provides a good indication of whether the condition will be respected in the final solution. 
If $\mathcal{M}^2$ is positive definite, we proceed to use this configuration as a seed for the evolutionary algorithm.

Note that the condition~\eqref{bubble equation AdS} for scaling solutions, and the scaling limit $r_{ab}\to\lambda r_{ab}$, correspond to ``zooming in'' to the core of the solutions, such that the asymptotics become AdS$_2$ fibered over $S^3$; for a general discussion, see~\cite{Bena:2018bbd}.
We wish to ``undo'' this limit and construct solutions with $\mathbb{R}^{4,1}$ asymptotics. This requires restoring the inhomogeneous terms in the bubble equations~\eqref{bubble equation original}. 

In the second step of the method, the evolutionary algorithm modifies the positions of the centers to construct numerical solutions to the full inhomogeneous bubble equations~\eqref{bubble equation original}, as we will describe in Section~\ref{sec: evolutionary algorithm}.

\subsection{The Bayesian optimization algorithm}\label{sec: BO}

We now describe the Bayesian optimization algorithm that we use to implement the first part of our method. 
As described in the previous section, we begin by specifying 
the positions 
of the centers $\vec r_a$ and the coefficients $q_a$, $l_0^i$, $k^{0,1}_a$ and $k^{2}_{0}$, considering them to be independent variables, and solving the homogeneous bubble equations~\eqref{bubble equation AdS} for $k^{2}_{\alpha}$ (recall $\alpha= 1,...,n-1$).

We wish to impose large charges, $Q_1,Q_5>\bar Q$, for some appropriate $\bar Q$. To do so, in principle one could attempt to maximise the first two expressions in Eq.~\eqref{global charges} as functions of both $k^{2}_{0}$ and the full set of $k^{0,1}_a$. However in practice, it is neither computationally efficient, nor necessary, to maximize $Q_1, Q_5$  with respect to a substantial fraction of the flux parameters $k^{0,1}_a$ --- it suffices to focus on the flux parameters of a small number of the centers.
We thus introduce a hyperparameter $n_{\textsc{b}}$ and maximize $Q_1, Q_5$ only with respect to the flux parameters of $n_{\textsc{b}}$ of the centers. In our examples, it will suffice to take $n_{\textsc{b}}$ to be equal to 1 or 2. 

The centers whose flux parameters are dependent variables of the optimization process will be labelled with the index $\bar a = 0,\cdots, n_{\textsc{b}}-1$, and the remainder will be labelled with the index $\dot a\,=\, n_{\textsc{b}}, \cdots, n-1 $. In other words, we will fix the value of the flux parameters $k^{0,1}_{\dot a}$ when initializing the algorithm, and we then maximize the value of the global charges with respect to $k^2_0$ and $k_{\bar a}^{0,1}$.

In order to maximize the value of the global charges we use a Bayesian optimization algorithm. The reason for this choice is that the global charges $Q_1, Q_5$ in~\eqref{global charges} are computationally expensive to evaluate: for any given trial configuration, one must first solve the homogeneous bubble equations for $k^2_{\alpha}$, and then use the result to compute $Q_1, Q_5$.

\vspace{2mm}
\noindent
{\bf Bayesian Optimization}

Let us now provide an intuitive description of how the Bayesian optimization algorithm works.  Bayesian optimization (BO) is an approach to find the global maximum (or minimum) of a ``black-box'' function, called the objective function. By black-box function, we mean either a function over which we have no analytic control (for example a stochastic function), or a function that is computationally expensive to evaluate, as in the case at hand. 
This means that we do not have a global knowledge of the function, i.e.~we do not know its value on every point of the domain, however we have the freedom to evaluate it on a finite set points.

In this context, Bayesian optimization algorithm is a strategy to obtain the maximum of such functions that works better than a random search. It works as follows (for a review, see e.g.~\cite{Frazier2018ATO}).  First, a Gaussian process prior is placed on the objective function. Then, the objective function is evaluated on a set of points $[x_1, \cdots, x_{\mathbf{n_0}}]$ of the domain. At this stage, the data $\{[x_1, \cdots, x_{\mathbf{n_0}}], [f(x_1), \cdots, f(x_{\mathbf{n_0}})]\}$ represent all our knowledge on the objective function. Of course, there are infinitely many functions whose value is $[f(x_1), \cdots, f(x_{\mathbf{n_0}})]\}$ when evaluated on $[x_1, \cdots, x_{\mathbf{n_0}}]$, but, by assuming that the objective function follows a Gaussian process model, we estimate that not all such functions are equally probable.

In the next step, we construct a ``surrogate'' function, which, among the infinite functions that have the same value of the objective on the points $[x_1, \cdots, x_{\mathbf{n_0}}]$, has the highest probability of representing the objective\footnote{This is why this optimization algorithm is called Bayesian: given the knowledge on the objective, the prior is used to obtain a posterior, which in this case is the surrogate function.}: as such, it is our best estimate of the objective based on the knowledge we have so far, and has the advantage of being much quicker to evaluate.

The last ingredient of the algorithm is the acquisition function, also known as acquisition strategy. By evaluating the {\it surrogate} function on a finite set of points of the domain, it chooses the next sampling point of the objective (i.e. the point of the domain that is more likely to pay off when the objective is evaluated on it), according to some specified strategy.  There are many different possible acquisition strategies; see e.g.~\cite{Frazier2018ATO}.

By evaluating the objective function on this new point, we increase the knowledge we have on the objective. Thus, after each additional sampling point the surrogate function is updated, and the acquisition function is again used to choose the next sampling point, iterating until an acceptably good approximate maximum (in our case, $Q_1, Q_5 > \bar Q$) of the objective is found, or a previously set computational limit is reached (in our case, a maximum number of iterations $N$).

\vspace{2mm}
\noindent
{\bf Implementation}

We now explain this first part of our algorithm in more detail.
We initialize the independent variables as follows. The $q_a$ must all sum to 1. If $n$ is odd, we use alternating values $\pm 1$, starting and ending with 1. If $n$ is even, the first $n-1$ use alternating values $\pm 1$, starting and ending with $-1$, while $q_n=2$:
\begin{equation}
   q_a = 
    \begin{cases}
    (1,-1,1,\ldots,-1,1) \qquad\quad  n\text{ odd}\\
    (-1,1,-1,\ldots,-1,2) \qquad~ n\text{ even}
    \end{cases}\,;
    \qquad\quad
    l_0^i=1\;\;~\forall\,i\,.
\end{equation}
 
Furthermore, we choose the position of the $n$ centers so that they lie inside a cube with edge length equal to two. We set up the coordinate system in such a way that the first center is at the origin, the second is at $y=0$, $z=0$ and the third center is at $z=0$, so that we have a total of $3n-6$ free coordinates. We sample the remaining non-zero $r^i_{a}$ from the following  uniform distribution:
\begin{equation}\label{r uniform distribution}
    r^i_{a}\,\sim\, U\big({}-1,1\big) \,.
\end{equation}
In practice, this sampling is obtained from a discrete distribution whose step-size is controlled by the hyperparameter~\texttt{prec\_pos}.
Similarly, the coefficients $k_{\dot a}^{0,1}$ are sampled with the discrete uniform distribution $U\big(-10^\texttt{prec\_k}, 10^\texttt{prec\_k}\big)$, with step-size equal to 1. In the following we will set the parameter~$\texttt{prec\_k}=2$.

Having initialized all the parameters we are not optimizing over, we now maximize the objective function
\begin{equation}
    f_{\textsc{obj}}\,=\, \min\big(Q_1,Q_5\big)\,,
\end{equation}
as a function of the variables $\{k^{0,1}_{\bar a}, k^2_0 \}$, which (having set $\texttt{prec\_k}=2$) we also allow to take integer values in $[-100,100]^{\otimes (2n_{\textsc{b}}+1)}$.
The evaluation of $f_{\textsc{obj}}$ works as follows. We first solve the homogeneous bubble equations~\eqref{bubble equation AdS}. Next, we round the flux parameters $k^2_\alpha\rightarrow \tilde k^2_\alpha$. Last, we use \eqref{global charges} to compute $Q_1, Q_5$ and select the minimum of the two.

We evaluate the objective function on $\mathbf{n_0}$ points in the $(2n_{\textsc{b}}+1)$-dimensional space of $\{k^{0,1}_{\bar a}, k^2_0 \}$ sampled from the discrete uniform distribution $U\big(-10^2, 10^2\big)^{\otimes (2n_{\textsc{b}}+1)}$. 
Assuming a Gaussian process prior as described above, we use knowledge of $f_{\textsc{obj}}$ evaluated at this set of points to generate the surrogate function. We use $\mathbf{n_0}=200$.

We then evaluate the surrogate function on a much higher number (of order $10^4$) of randomly sampled points, with discrete uniform distribution $U\big(-10^2, 10^2\big)^{\otimes (2n_{\textsc{b}}+1)}$ and step-size 1. We use an acquisition function based on the Probability of Improvement method (see e.g.~\cite{Agnihotri2020}) to choose the next point of the domain that is most worth evaluating with the objective function. The knowledge of $f_{\textsc{obj}}$ at this new point is then used to update the surrogate function. This process is iterated until a point $\{k^{0,1}_{\bar a}, k^2_0 \}$ such that $f_{\textsc{obj}}\big(\{k^{0,1}_{\bar a}, k^2_0 \}\big)>\bar Q$ is found, or a previously set computational limit (the maximum number of iterations $N$) is reached. This procedure is summarized in Algorithm~\ref{alg:BO}.

\begin{algorithm}
\caption{BO algorithm}\label{alg:BO}
\begin{algorithmic}
\Function{$f_{\textsc{obj}}$}{$\{k^{0,1}_{\bar a}, k^2_0 \}$}
\State $k^2_{\alpha}\gets$ Solve the homogeneous bubble equations~\eqref{bubble equation AdS}
\State $\tilde k^2_\alpha \gets$ Round $k^2_{\alpha}$
\State Compute $Q_1, Q_5$ via \eqref{global charges}
\State\Return $\min\big(Q_1,Q_5\big)$
\EndFunction
\State Assume Gaussian process prior
\State Evaluate $f_{\textsc{obj}}$ on $\mathbf{n_0}$ points $\{k^{0,1}_{\bar a}, k^2_0 \}$ sampled with $U\big(-10^2, 10^2\big)^{\otimes (2n_{\textsc{b}}+1)}$
\State $Q_{\text{max}}\gets$ The maximum output of $f_{\textsc{obj}}$ found so far
\State Generate the surrogate function using the available data
\While{$n < N$ or $Q_{\text{max}}<\bar Q$}
\State Let $\{k^{0,1}_{\bar a}, k^2_0 \}_n$ be the point returned by the acquisition function
\State Evaluate $f_{\textsc{obj}}\big(\{k^{0,1}_{\bar a}, k^2_0 \}_n\big)$
\State $Q_{\text{max}}\gets$ if $f_{\textsc{obj}}\big(\{k^{0,1}_{\bar a}, k^2_0 \}_n\big)>Q_{\text{max}}$, update $Q_{\text{max}}$
\State Update the surrogate function with the new data
\State $n$++
\EndWhile
\end{algorithmic}
\end{algorithm}

While this algorithm is not guaranteed to find a solution, we found that in practice this approach is much more successful than a random search.

After a successful run of the Bayesian optimization algorithm, we impose the two conditions described at the end of Sec.~\ref{sec:algo-overview}, to be confident that a nearby scaling solution exists and that there are no CTCs in the configuration at this point. If and only if these two tests are passed, we proceed to the evolutionary algorithm.

\subsection{The evolutionary algorithm}\label{sec: evolutionary algorithm}

A successful run of the Bayesian optimization algorithm outputs an approximate solution to the homogeneous bubble equations~\eqref{bubble equation AdS} with the desired characteristics. The approximate nature of this solution is due to the rounding of the flux parameters $k_{\alpha}^2$. Our task is now to obtain a numerical solution of the inhomogeneous bubble equations~\eqref{bubble equation original} by moving the positions of the centers in the $\mathbb{R}^3$ base space.

We shall do so by using an evolutionary algorithm (EA), which is an optimization algorithm inspired by Darwin's theory of evolution. The starting point is a {\it population}, i.e.~a set of {\it individuals} that are approximate solutions to the problem we wish to solve. The properties of each individual are called {\it genes}. We measure how good an approximate solution is via the {\it fitness} function, which is the function we want to maximize. The fittest individuals are selected to reproduce, by passing some of their genes to their offspring and in the reproduction process some mutations are implemented, i.e.~random modifications of the genes of the offspring. Then the offspring take the place of the less fit individuals, which die, such that the population size is constant. This process is iterated until a sufficiently good solution is found, or a previously set computational limit is reached.

For the case at hand, each individual is an approximate solution to the bubble equations~\eqref{bubble equation original}. An individual's genes are (a subset of) the positions of the centers together with a set of {\it strategy parameters}, to be described momentarily.

After implementing the translational and rotational symmetry, the number of free coordinates is $3n-6$, while the number of independent bubble equations is $n-1$. 
Implementing a genetic algorithm on all the $3n-6$ coordinates would be computationally expensive. Thus, we select a subset of the coordinates to be fixed to the values of the BO algorithm output, $\bar{r}_a^i$ of Eq.~\eqref{r rescaled}. 
We observed that fixing approximately one coordinate on each of the last $n-3$ centers provides a good balance between effectiveness and computational cost, and we shall give explicit examples of this in Section~\ref{results}. We shall denote by $d$ the number of degrees of freedom, i.e.~the number of unfixed coordinates over which we run the EA.

To describe the algorithm further, we introduce the multi-index $A=(a,i)$ combining the center label $a$ and the three Euclidean coordinates $i$. For the $p^{\mathrm{th}}$ individual in the population, we denote the set of coordinates to be varied by $r_A^{(p)}$. 
As noted above, we work directly with the positions of the centers as genes, so the distances between the centers are always well-defined.

In the reproduction process, random mutations occur. The magnitude of the mutations is controlled by the set of strategy parameters. Each direction $r_A^{(p)}$ has an independent strategy parameter $\sigma^{(p)}_A$, which is a gene of the individual, and which also undergoes variation and selection itself. As we will describe in the following, we will initialize the positions of the individuals in the population by sampling from a Gaussian distribution with fixed standard deviation, which will be taken as the initial value of the strategy parameter for every individual and every direction.
The genes of the $p$-th individual in the population are then written as
\begin{equation}\label{individual genes}
\{r_A^{(p)},\sigma^{(p)}_A\}
\,, \qquad\quad A=(a,i)\,.
\end{equation}

All genes $\{r_A^{(p)},\sigma^{(p)}_A\}$ undergo variation and selection.
An individual whose genes are likely to survive in the evolutionary process will have a good set of positions $r_A^{(p)}$, that will be quantified by having high fitness, 
and a set of strategy parameters $\sigma^{(p)}_A$ that are likely to give rise to fit offspring, as will become clearer when we discuss reproduction and mutation. 

The evolutionary mechanism is iterated over a certain number of generations controlled by the hyperparameter $\texttt{generations}$; in each new generation a number of offspring is generated, specified by the hyperparameter $\texttt{offspring\_per\_generation}$.
The optimal values of these hyperparameters depend on the number of degrees of freedom of the problem, i.e.~the number of centers of the configuration, as we shall see in examples.

\vspace{2mm}
\noindent
{\bf Fitness function}

We now define the fitness function, i.e.~the function the EA algorithm seeks to maximize.
We seek solutions to the inhomogeneous bubble equations~\eqref{bubble equation original}, so we use these to construct the fitness function.
By rearranging the bubble equations, we write:
\begin{equation}\label{fitness 1}
    \sum_{a\neq b}\frac{q_a q_b}{r_{ab}}\Pi^0_{ab}\Big(\Pi^1_{ab}\Pi^2_{ab}-\frac{1}{2g^2} \mathbb{T}_{ab}\Big)-\sum_{b,i} q_aq_bl^i_0\Pi^i_{ab}\,=\,\epsilon_a\,,
\end{equation}
where if $\epsilon_a\,=\,0~\,\forall a\,$, then the solution is exact.
We seek configurations $\{r_A\}$ that minimize the errors associated to all the bubble equations. We do so by instructing the algorithm to minimize $\sum_a |\epsilon_a|$. That is, we define the algorithm's fitness function to be
\begin{equation}\label{fit fun}
    f(r_{A})\,=\,\frac{1}{\sum_a |\epsilon_a|}\,.
\end{equation}
It will also be useful to define the following inverse fitness function:
\begin{equation}\label{inv fit fun}
    f_{\text{inv}}(r_{A})\,=\,\sum_a |\epsilon_a|\,.
 \end{equation}

Note that since the sum of the bubble equations is zero by construction, the inverse fitness function $f_{\text{inv}}$ typically over-states the error of a configuration, particularly as the number of centers grows larger. A more accurate measure is the largest absolute value of the errors of the individual equations. So, for later use when discussing our results, we also define the following maximum-error fitness and inverse fitness functions,

\begin{equation}\label{fit fun max}
    \tilde{f}(r_{A})\,=\,\frac{1}{\max_a |\epsilon_a|}\;,
\end{equation}
and
\begin{equation}\label{inv fit fun max}
\tilde{f}_{\text{inv}}(r_{A})\,=\,{\max_a |\epsilon_a|}\;.
\end{equation}

\vspace{0.5mm}

\noindent
{\bf Population initialization}

\vspace{0.5mm}
We initialize a population of \texttt{pop\_size} individuals by starting with the configuration of centers obtained in~\eqref{r rescaled}, and adding to it a random variable $\delta r_A^{(p)}$ sampled from the Gaussian distribution $\mathcal N(0,\texttt{var\_pos})$ with mean $0$ and standard deviation \texttt{var\_pos}, where \texttt{var\_pos} is a hyperparameter of the algorithm:
\begin{equation}\label{new individual}
    r_A^{(p)}\,=\,\bar{r}_A+\delta r_A^{(p)}\,\qquad\text{for}\; p=1,\cdots,\texttt{pop\_size}\,.
\end{equation}
The optimal value of \texttt{var\_pos} depends on the number of centers. If \texttt{var\_pos} is too small, we generate a population that is too similar to the seed solution; if \texttt{var\_pos} is too large, we obtain a large number of unfit individuals in the initial population.
It is not practical to compute optimal value of \texttt{var\_pos} directly from the bubble equations Eq.~\eqref{bubble equation original}, so we optimize it via a simple grid search. Concretely, we examine the populations generated by a set of candidate values of \texttt{var\_pos}, and use the population fitness to select the optimal \texttt{var\_pos}.

Once an individual is generated, we implement the opposition-based technique \cite{tizhoosh2005opposition,rahnamayan2006opposition}, i.e. we compare this individual with the one obtained through reflection symmetry with respect to the initial configuration~\eqref{r rescaled}, and keep the one which has the highest fitness. 
In other words, given the individual with position $r^{(p)}_A$, we consider the individual with position $r'^{(p)}_A$ given by:
\begin{equation}\label{opposition base individual}
   r'{}_A^{(p)}=2\bar r_A-r^{(p)}_A\,,
\end{equation}
and add to the population (only) the fitter of the two individuals.
We initialize the strategy parameter as $\sigma^{(p)}_A\,=\,\texttt{var\_pos}$ for all $p,A$.

\vspace{3mm}

\noindent
{\bf Selection}

\vspace{0.5mm}

The selection mechanism of the evolutionary algorithm dictates which individuals pass their genes to the offspring, and which individuals are replaced in the new generation. Recall that the number of new offspring per generation is a hyperparameter of the algorithm. 
To produce an offspring, the algorithm selects two parents that reproduce and one individual that will be replaced.
This selection process occurs probabilistically,  favouring potential parents with highest fitness, and where those with highest inverse fitness are most likely to die off. 

Our algorithm implements two different selection mechanisms, amongst which the user can choose. The two methods are~(see e.g.~\cite{eiben2003introduction} for more details):
\begin{itemize}
    \item Fitness proportional selection. The probability of an individual to be chosen as a parent is:
    \begin{equation}\label{Fitness proportional selection}
        P^{(p)}=\frac{f({r^{(p)}_A})}{\sum_p f({r^{(p)}_A})}\,,
    \end{equation}
    where $f$ is the fitness function~\eqref{fit fun}.
    The probability of an individual to die off is governed by the same equation, with the fitness function replaced by the inverse fitness function, defined in Eq.~\eqref{inv fit fun}.
    
    \item Sigma scaling. The probability of an individual to be chosen as parent is  given by a modified version of Eq.~\eqref{Fitness proportional selection}, with the fitness function being replaced by the following auxiliary fitness function $f'$, which involves a shift controlled by the mean $\bar f$ and standard deviation $\sigma_f$ of the fitnesses of the population:
    \begin{equation}\label{sigma scaling}
    f'(r^{(p)}_A)=\max\Big(f\big(r^{(p)}_A\big)-\big(\bar f-c\;\! \sigma_f\big),0\Big)\,,
    \end{equation}
    where $c$ is a constant which is usually set to 2. Similarly, we also apply this method to the selection of the individual that dies by replacing the fitness function with the inverse fitness function.
\end{itemize}
The first method is less computationally expensive, however it can be less effective due to the following disadvantage. It is sensitive to adding a constant shift to all $f$, and depending on this shift there can be too much or too little selection pressure. For instance, if there is too much selection pressure, the best individuals tend to dominate the population very quickly, which can lead to premature convergence. 

By contrast, the second method is more computationally expensive, but tends to produce an appropriate amount of selection pressure.

\vspace{2mm}
\noindent
{\bf Reproduction and mutation}

\vspace{0.5mm}
Once two parents $\{r^{(1)}_{A}, \sigma_A^{(1)}\}$ and $\{r^{(2)}_{A}, \sigma_A^{(2)}\}$ are selected, their offspring $\{ r_{A},  \sigma_A\}$ is generated as follows. First, separately for each gene, i.e.~for each element of the multi-index $A$, we assign equal probability to one of the following three reproduction methods to occur. Before possible mutation, this gene will be set equal to the gene of a parent, or the average of the two parental genes:
\begin{equation}\label{reproduction}
    \begin{aligned}
    1)&\qquad \{ r_{A},  \sigma_A\}
            \,=\,\{r^{(1)}_{A}, \sigma_A^{(1)}\}\,;\\
    2)&\qquad \{ r_{A},  \sigma_A\}
            \,=\,\{r^{(2)}_{A}, \sigma_A^{(2)}\}\,;\\
    3)&\qquad \{r_{A}, \sigma_A\}
            \,=\,\Big\{\frac{r^{(1)}_{A}+r^{(2)}_{A}}{2}, \frac{\sigma_A^{(1)}+\sigma_A^{(2)}}{2}\Big\}\,.\\
    \end{aligned}
\end{equation}

Next, we implement a mutation mechanism, which can enable the population to escape from local minima of the fitness function.
Recall that each offspring has a total of $d$ genes ($A=1,\ldots,d$). For each gene of an offspring, we assign a probability of order $2/d$ for the gene to mutate away from the value assigned in \eqref{reproduction}. So on average, approximately two genes of each offspring will mutate.

It is desirable to build in the flexibility for mutations in different genes to have different strengths.
We therefore implement uncorrelated mutation, with different step sizes for different genes~(see e.g.~\cite{eiben2003introduction}). 
If a gene is selected for mutation,
first $\sigma_A$ mutates, then the mutated $\sigma'_A$ sets the scale for the mutation of the position $r_A$.
The mutation of $\sigma_A$ is controlled by two Gaussian random variables: $\delta \sigma$ is sampled only once for each offspring, while $\delta\sigma_A$ is sampled separately for each gene. Both are sampled from the Gaussian distribution $\mathcal N(0,1)$. The mutation strength of $\sigma_A$ is controlled by the parameters $\tau'=1/\sqrt{d}$ and $\tau=1/\sqrt{2\sqrt{d}}$ via:
\begin{equation}
\label{eq:mut-1}
    \sigma_A
    \,\to\,\sigma_A'\,=\,\sigma_A \exp{\Big(\tau'\, \delta \sigma +\tau\, \delta \sigma_A\Big)}\,,
\end{equation}
where following~\cite{eiben2003introduction}, we set
$\tau'=1/\sqrt{d}$ and $\tau=1/\sqrt{2\sqrt{d}}$.
The way this treats different directions differently is as follows.  The mutation $\exp(\tau'\, \delta \sigma)$ is common to all directions and allows for an overall change of the mutation step-size. By contrast, the term $\exp(\tau\, \delta \sigma_A)$ introduces different mutation strengths in different directions.

Once $\sigma_A$ has mutated to $\sigma'_A$, the position $r_A$ mutates with a Gaussian random variable $\delta r_A$ drawn from $\mathcal N(0,1)$, with mutation strength set by the new $\sigma'_A$:
\begin{equation}
\label{eq:mut-2}
  r_A\,\to\,r_A'\,=\,r_A+\sigma'_A \, \delta r_A\,.
\end{equation}

As the fitness of the population improves, and the algorithm explores narrower regions of the phase space, we improve upon the mutation mechanism 
\eqref{eq:mut-1} to achieve the following set of goals. 
The mutation should enable an appropriately fine exploration of the narrower region, while not becoming too small in magnitude. Separately, the mutation should be able to jump outside local minima. 
To enable this, we introduce a hyperparameter $\texttt{generation\_update}$. After $\texttt{generation\_update}$ generations, we update the strategy parameter of the individuals according to the following prescription. After this update, the algorithm returns to the mutation \eqref{eq:mut-1} for the next $\texttt{generation\_update}$ generations. 
The algorithm contains two different methods to update the strategy parameter, among which the user can choose. These are:
\begin{itemize}
    \item Random update. We introduce the hyperparameters $\texttt{percentage\_random\_update}$ and $\texttt{factor\_random\_update}$,  randomly select  $(\texttt{percentage\_random\_update})\%$ of the individuals and rescale $\sigma_A\rightarrow \sigma_A/\texttt{factor\_random\_update}$, while rescaling the strategy parameter of the remaining individuals as $\sigma_A\rightarrow( \texttt{factor\_random\_update})\, \sigma_A$ .
    \item Variance update.  We update the strategy parameter according to the variance of the positions, as follows. For each $A$, we define the average position
    \begin{equation}
        \hat r_A\,=\,\frac{1}{\texttt{pop\_size}}\sum_q r_A^{(q)}\,,
    \end{equation}
    and reset the strategy parameters in direction $A$ of all individuals to have the same value,
    \begin{equation}
        \label{var update strategy}\sigma_A^{(p)}\,\rightarrow\,\sigma'_A{}^{\!\!(p)}\,=\, \sqrt{\frac{1}{\texttt{pop\_size}}\sum_q\Big(r_A^{(q)}-\hat r_A\Big)^2}\, \qquad~~~ \forall p \,.
    \end{equation}
    The variance update method works as follows. If, for instance, the whole population is concentrated in a particular region of parameter space, it tends to enable finer exploration of the local region. By contrast, if for instance the population is concentrated in two or more  separate regions, the variance update tends to 
    enable the population to 
    explore a larger region of the parameter space. 
    
\end{itemize}
We performed several runs with both Random and Variance updates. The Variance update has the advantage that it does not depend on the number of centers, while in the Random update we are introducing two new hyperparameters that could in principle be optimized over, depending for instance on the number of centers. Of the two methods, typically the Variance update is more computationally expensive, and typically the performance of the algorithm is better than or comparable to the Random update, depending on the other parameters of the configuration.

For the sake of clarity, let us illustrate an example of how the variance update method works. Suppose we set the evolutionary algorithm to run for $1000$ generations, and that we specify $\texttt{generation\_update}=100$. The evolution of the strategy parameters $\sigma_A$ will work as follows. Up to generation 99, the $\sigma_A$ of each individual will evolve with random mutations via~Eq.~\eqref{eq:mut-1}. Then, at generation $100$, the strategy parameters $\sigma_A$ of all individuals in the population will be reset according to Eq.~\eqref{var update strategy}. Next, for generations $101$ to $199$ we return to the random mutations described in Eq.~\eqref{eq:mut-1}. Then, at generation $200$ we again reset the $\sigma_A$ according to Eq.~\eqref{var update strategy}, and the pattern continues until the end.

\section{Results}\label{results}

In this section we present two explicit examples of numerical scaling solutions obtained with the algorithm described above, which have five and seven centers respectively. In both of these examples we use the fitness proportional selection method and the variance update mechanism described in Section~\ref{sec: evolutionary algorithm}. In the examples that we present, the non-Abelian coupling constant $g$ will be set equal to $1$. We also comment on the performance of the algorithm.

\subsection{A five-center scaling configuration}\label{sec: 5 centers example}

As a first application of our method, we provide an example of a five-center scaling configuration. 
We first optimize the hyperparameters of the evolutionary algorithm for a five-center configuration. We do so with a grid search and obtain the values reported in Table~\ref{tab: Hyperparameter 5 cent}.

\vspace{1mm}
\begin{table}[h]
\centering
\begin{tabular}{l*{6}{c}r}
 $\texttt{pop\_size}$ &$\texttt{offspring\_per\_generation}$&$\texttt{generations}$&$\texttt{var\_pos}$&$\texttt{generation\_update}$  \\
\hline

2000 & 50 & 16500 & $10^{-4}$ &666  \\
\end{tabular}
\label{tab: Hyperparameter 5 cent}
\caption{Hyperparameters of the algorithm, optimized for five-center configurations.}
\end{table}

We then follow the procedure described in Section~\ref{sec: BO}. We use the Bayesian optimization algorithm to obtain initial positions and flux parameters, recorded in Table~\ref{tab: 5 centers}, that give a configuration with global charges $Q_1\simeq 2938$  and $Q_5\simeq 2016$.

\vspace{1mm}
\begin{table}[h]
\centering
\begin{tabular}{l*{6}{c}r}
& $1^{\text{st}}$ & $2^{\text{nd}}$  & $3^{\text{rd}}$ &$4^{\text{th}}$ & $5^{\text{th}}$ \\
\hline
$x$     & $0$ & \underline{0.3314} & \underline{0.7491} & $-0.6923$ & \underline{0.4644}   \\
$y$     & 0 & 0 & \underline{0.5648} & $-0.684$ & \underline{0.4799} \\
$z$     & 0 & 0 & 0 & \underline{$-0.0792$} &  0.549   \\
$q$     & 1 & 1 & $-1$ & $-1$ &  1   \\
$k^0$     & 17 & $-18$ & 63 & 47 &  29   \\
$k^1$     & $-61$ & 66 & 25 & 72 & 80   \\
$k^2$     & 60 &- &-  &-  &-    \\
\end{tabular}
\label{tab: 5 centers}
\caption{Input parameters of the configuration. For ease of notation, the rescaling in Eq.~\eqref{r rescaled} with $\lambda\,=\,10^{-5}$ is understood: in the first three rows of this table the coordinates of the centers are in units of $10^{-5}$, i.e.~$r^1_2 = 0.3314 \times 10^{-5}$. The underlined coordinates are those that are genes of the EA, i.e.~the coordinates over which the evolution process occurs.}
\end{table}

By solving the homogeneous bubble equations~\eqref{bubble equation AdS}, we obtain the $(n-1)$ remaining  $k^2_\alpha$ parameters. After rounding to a precision of $\texttt{k\_rounding}\,=\,10^{-4}$, we report their values in Table~\ref{tab: 5 centers fluxes}.

\begin{table}[!ht]
\centering
\begin{tabular}{l*{6}{c}r}
 & $1^{\text{st}}$ & $2^{\text{nd}}$  & $3^{\text{rd}}$ &$4^{\text{th}}$ & $5^{\text{th}}$   \\
\hline

$k^2$   &-  & 56.0815 & $-48.5265$ & $-51.1402$  & 47.9087 \\
\end{tabular}
\label{tab: 5 centers fluxes}
\caption{Solution of the homogeneous bubble equations, with the input parameters given in Table~\ref{tab: 5 centers}, after rounding.}
\end{table}

The parameters in Tables~\ref{tab: 5 centers} and~\ref{tab: 5 centers fluxes} define a numerical solution to the inhomogeneous bubble equations~\eqref{bubble equation original} with fitness $f\big(\bar r^i_a \big)\approx 4.26$. The configuration respects the condition regarding the possibility of a nearby scaling solution discussed around Eq.~\eqref{flux signs}. Moreover, the matrix $\mathcal{M}^2$ is positive-definite. We therefore  proceed to use the configuration as a seed configuration in the evolutionary algorithm. 
The genes of the evolutionary algorithm are the subset of that are coordinates underlined in Table~\ref{tab: 5 centers}.

\begin{figure}[t]
    \centering
    \includegraphics[width=14cm]{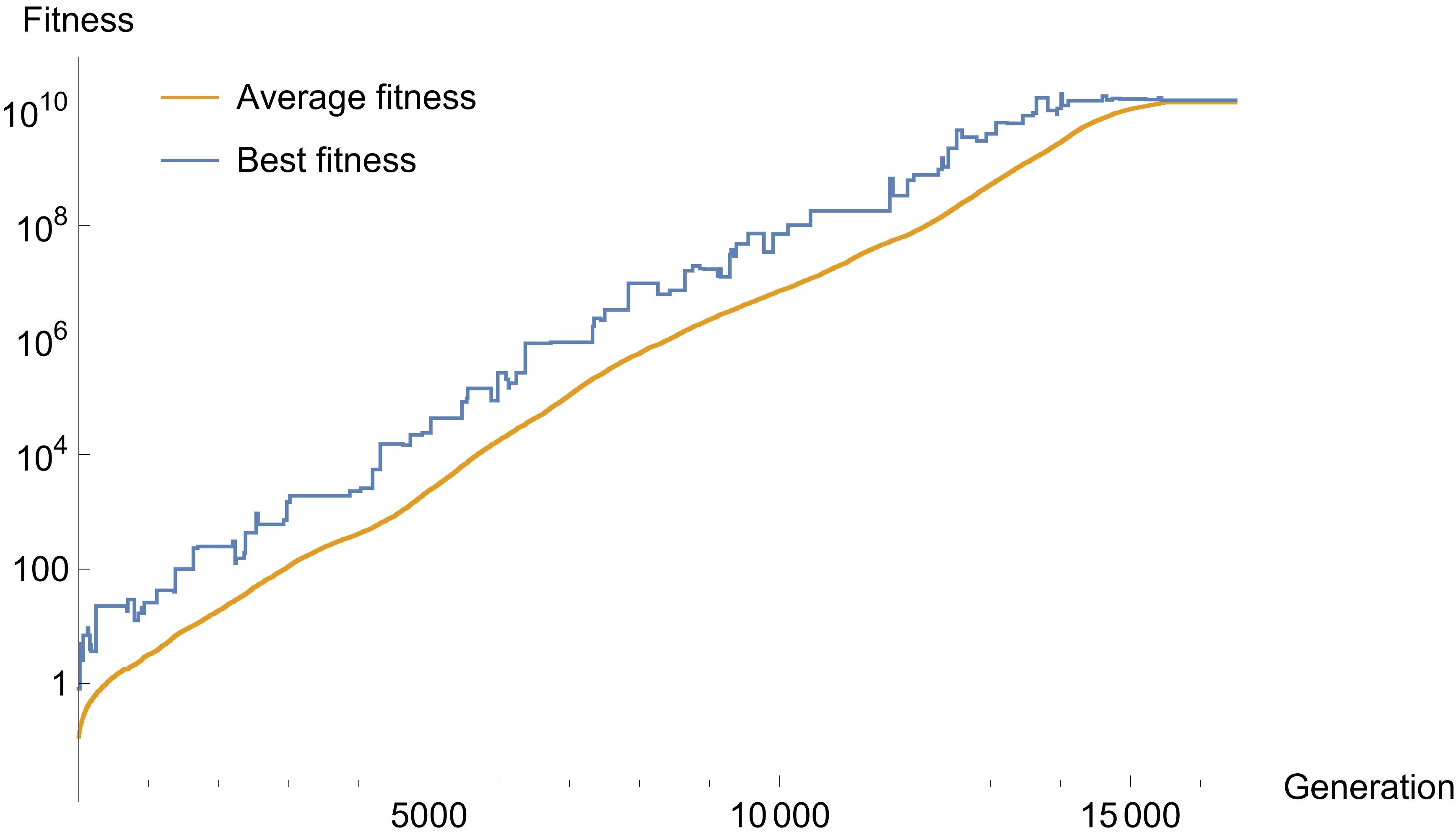}
    \caption{Fitness of the fittest five-center configuration, and average fitness of the population over the generations. The final plateau occurs at machine precision, as discussed in the text. The maximum fitness obtained is $\tilde{f}\simeq 2 \times 10^{10}$, corresponding to a dimensionless relative error of $1 \times 10^{-15}$.}
    \label{fig:5 cent}
\end{figure}

We plot in Figure~\ref{fig:5 cent} how the fitness of the fittest individual and the average fitness of the population change over the generations in a representative run of the algorithm. By starting with a seed solution with fitness of order $1$ we obtain, after around 14000 generations, a numerical solution with fitness $\tilde{f}\simeq 2\times 10^{10}$. This means that the EA generated a numerical solution that solves each bubble equation with a precision of at least $5 \times 10^{-11}$.

Note that these fitness/error figures are dimensionful, and depend on our choice of units. A useful dimensionless relative error can be defined by comparing the error in the bubble equations to the largest summand on the left-hand side of the bubble equations. For this five-center configuration, the largest summand has absolute value approximately $5\times 10^4$. 
The ratio of the largest error in the bubble equations, $\tilde{f}_{\text{inv}}(r_{A})\,=\,{\max_a |\epsilon_a|}$, to this largest summand, is therefore approximately $1\times 10^{-15}$. 
This is a typical best-case value at which the dimensionless relative error of the algorithm saturates, reflecting the fact that we work to machine precision.

As a side point, we observe that the fitness of the fittest individual in the population does not increase monotonically. The main reason is that occasionally the fittest individual is selected to die; this is part of the random nature of the algorithm, and is one way that the algorithm can escape local minima. A more minor reason is that the algorithm maximizes the sum-error fitness $f$ defined in~\eqref{fit fun}, while we have plotted the max-error fitness $\tilde{f}$ defined in~\eqref{fit fun max}, being a more accurate measure of the fitness.

\begin{table}[!ht]
\centering
\begin{tabular}{l*{6}{c}r}
     & $1^{\text{st}}$ & $2^{\text{nd}}$  & $3^{\text{rd}}$ &$4^{\text{th}}$ & $5^{\text{th}}$ \\
\hline
$x$     & $0$ &0.3314045307889   & 0.7491026359717&-0.6923 &0.4644170857933   \\
$y$     & 0 & 0 &0.5648088034751 & -0.6840217263604& 0.4799 \\
$z$     & 0 & 0 & 0 &-0.07910622500586 &  0.549   \\
\end{tabular}
\label{tab: 5 centers output}
\caption{Output of the evolutionary algorithm with highest fitness, for the five-center configuration. 
Similarly to those in Table~\ref{tab: 5 centers}, these coordinates are in units of $10^{-5}$.}
\end{table}

The associated matrix $\mathcal{M}^2$ is positive-definite, thus we can have confidence that the geometry does not contain any CTCs.
We report in Table~\ref{tab: 5 centers output} the coordinates of the centers of this numerical solution.

We compute the global charges of this solution using Eq.~\eqref{global charges}, obtaining: 
\begin{equation}
\begin{aligned}
 &Q_1\,\simeq\,   2938\,,\qquad Q_5\,\simeq\,   2016\,,\qquad Q_P\,\simeq\,  2.998 \times 10^4\,,\\
 &J_L\,\simeq\,   4.090\times 10^5\,,\qquad J_R\,\simeq\,   0.001545\,,
 \end{aligned}
\end{equation}
where $J_R \equiv |\vec{J}_R|$. 
We note that $J_R \ll 1$ as expected.

\subsection{A seven-center scaling configuration}
We now present an example of a seven-center scaling configuration.
We report in Table~\ref{tab: Hyperparameter 7 cent} the hyperparameters of the evolutionary algorithm, optimized for seven-center configurations.

\vspace{1mm}
\begin{table}[h!]
\centering
\begin{tabular}{l*{6}{c}r} $\texttt{pop\_size}$ &$\texttt{offspring\_per\_generation}$&$\texttt{generations}$&$\texttt{var\_pos}$&$\texttt{generation\_update}$  \\
\hline

3000 & 100 & 18500 & $10^{-4}$ &666  \\
\end{tabular}
\label{tab: Hyperparameter 7 cent}
\caption{Hyperparameters of the algorithm, optimized for seven-center configurations.}
\end{table}

After running the Bayesian optimization algorithm, we obtain an initial configuration with global charges $Q_1\simeq 736$  and $Q_5\simeq 410$, which is described in Table~\ref{tab: 7 centers}. 

\vspace{1mm}
\begin{table}[h]
\centering
\begin{tabular}{l*{6}{c}r}
     & $1^{\text{st}}$ & $2^{\text{nd}}$  & $3^{\text{rd}}$ &$4^{\text{th}}$ & $5^{\text{th}}$ &$6^{\text{th}}$ &$7^{\text{th}}$   \\
\hline
$x$     & $0$ & \underline{0.8409}  & \underline{0.3858} &-0.0195 & \underline{0.2188} &\underline{-0.6853} &\underline{-0.6294}   \\
$y$     & 0 & 0 &\underline{-0.4476} & \underline{-0.8449} & -0.2569 & \underline{-0.82} &\underline{0.8284}\\
$z$     & 0 & 0 & 0 & \underline{-0.2854} &  \underline{-0.9864}   & \underline{-0.3303}&-0.9516\\
$q$     & 1 & -1 & 1 & -1 &  1 & -1 & 1   \\
$k^0$     & 6 & -38 &56 & 38 &  86 & 85 & 15  \\
$k^1$     & 99 & 57 & 39 & 30 & 48 & 37 & 72  \\
$k^2$     & 39 &- &-  &-  &- &-&-    \\
\end{tabular}
\label{tab: 7 centers}
\caption{Input parameters of the solution. 
Similarly to those in Table~\ref{tab: 5 centers}, the coordinates in the first three rows are in units of $10^{-5}$, and underlined coordinates are genes of the evolutionary algorithm.}
\end{table}

The homogeneous bubble equations~\eqref{bubble equation AdS} give the $n-1$ remaining flux parameters $k^2_\alpha$; we round them to a precision of $10^{-5}$, and report the result in Table~\ref{tab: 7 centers fluxes}.

\vspace{1mm}
\begin{table}[h!]
\centering
\begin{tabular}{l*{6}{c}r}
 & $1^{\text{st}}$ & $2^{\text{nd}}$  & $3^{\text{rd}}$ &$4^{\text{th}}$ & $5^{\text{th}}$ &$6^{\text{th}}$ &$7^{\text{th}}$   \\
\hline

$k^2$   &-  & -37.2187 & 38.5597& -38.568  &38.4874 &-38.6549 & 38.8499    \\
\end{tabular}
\label{tab: 7 centers fluxes}
\caption{Solution of the homogeneous bubble equations, with input parameters given in Table~\ref{tab: 7 centers},  after rounding.}
\end{table}
The coefficients in Tables~\ref{tab: 7 centers} and~\ref{tab: 7 centers fluxes} provide an approximate solution to the inhomogeneous bubble equations~\eqref{bubble equation original} with fitness $f\big(\bar r^i_a \big)\approx 3.21$.
As in the five-center example, 
only the underlined coordinates in Table~\ref{tab: 7 centers} are taken to be genes of the EA.

\begin{figure}[t!]
    \centering
    \includegraphics[width=14cm]{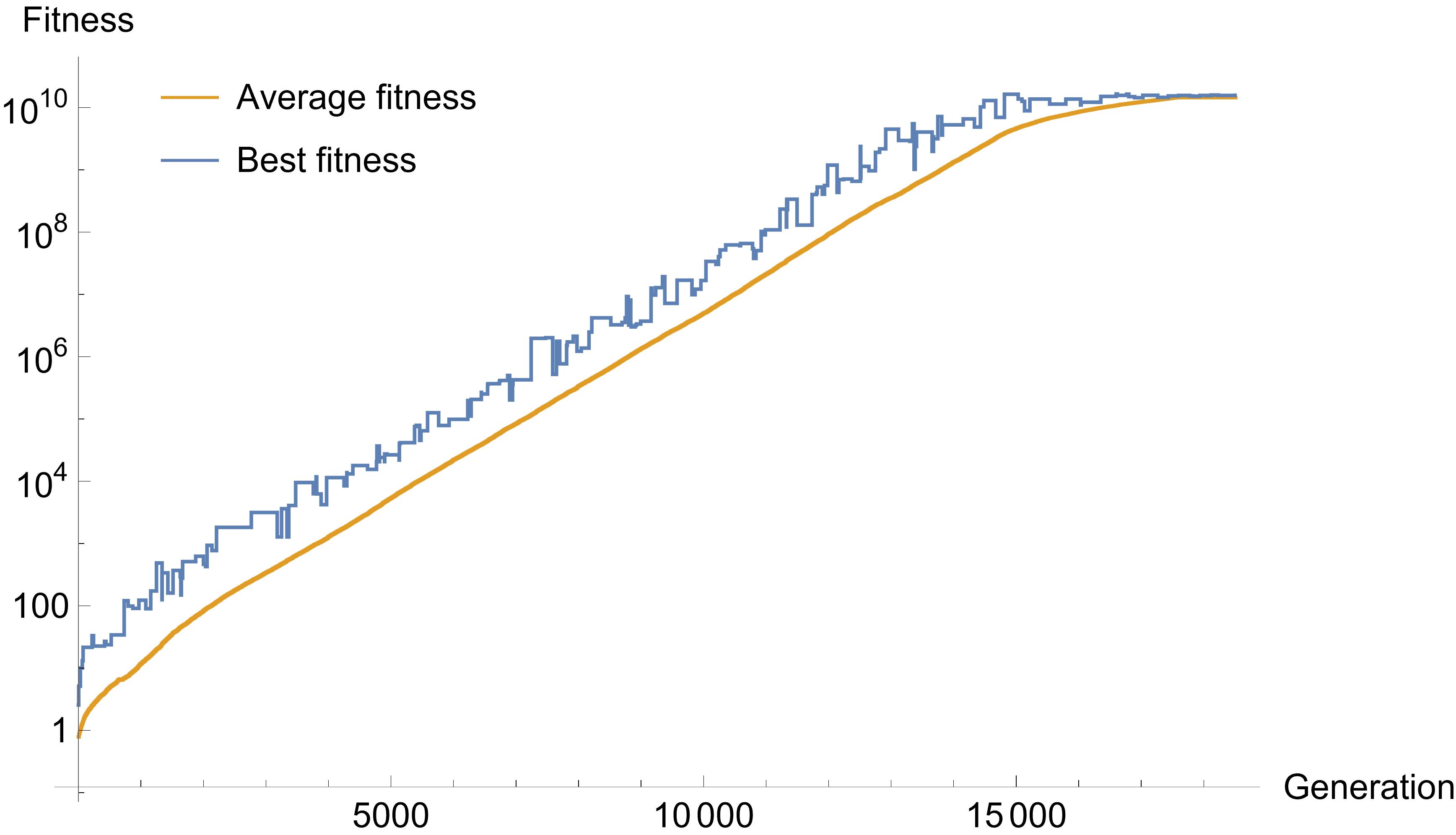}
    \caption{Fitness of the fittest seven-center configuration, and average fitness of the population over the generations. The final plateau again occurs at machine precision: the maximum fitness obtained is $\tilde{f}\simeq 1.7 \times 10^{10}$, corresponding to a dimensionless relative error of $5.7 \times 10^{-15}$.}
    \label{fig:7 cent}
\end{figure}

As depicted in Figure~\ref{fig:7 cent}, we obtain, after around 17000 generations, a numerical solution with fitness $\tilde{f}\simeq 1.7 \times 10^{10}$. 
For this configuration,
the absolute value of the largest term on the left-hand side of the bubble equations is $1.1\times 10^4$. So the dimensionless relative error defined in the previous subsection is approximately $5.7 \times 10^{-15}$, again reflecting the fact that we work to machine precision.

The coordinates of the centers of the fittest configuration obtained by the evolutionary algorithm are reported in Table~\ref{tab: 7 centers output}.

\vspace{2mm}
\begin{table}[h]
\centering
\begin{tabular}{l*{6}{c}r}
     & $1^{\text{st}}$ & $2^{\text{nd}}$  & $3^{\text{rd}}$ &$4^{\text{th}}$ & $5^{\text{th}}$ &$6^{\text{th}}$ &$7^{\text{th}}$   \\
\hline
$x$     & $0$ &0.84088499  &0.38578191  &0.0195 &0.21880766 &-0.68529849 &-0.62944182   \\
$y$     & 0 & 0 &-0.44759418  &-0.84485200   &-0.2569   & -0.82004941  &0.82839175 \\
$z$     & 0 & 0 & 0 &-0.28532553  &-0.98638729     &-0.33036911  & -0.9516\\
\end{tabular}
\label{tab: 7 centers output}
\caption{Output of the evolutionary algorithm with highest fitness, for the seven-center configuration. Again, coordinates are in units of $10^{-5}$.}
\end{table}
The matrix $\mathcal{M}^2$ is positive-definite, and thus we can have confidence that the configuration is free of CTCs.
Finally, we record the global charges of the solution: 
\begin{equation}
\begin{aligned}
 &Q_1\,\approx\,   736.3\,,\qquad Q_5\,\approx\,   410.0\,,\qquad Q_P\,\approx\,  8.887\times 10^4\,,\\
 &J_L\,\approx\,   1.625\times 10^5\,,\qquad J_R\,\approx\,   0.007047\,.
 \end{aligned}
\end{equation}
We again observe that $J_R\ll 1$ as expected.

\subsection{Performance of the algorithm}\label{sec: perfomance}

We now make some general comments regarding the algorithm's performance.
We have run the algorithm in detail for configurations of three, five, and seven centers. As the number of centers increases, naturally the runtime increases. This is primarily due to two factors. First, the number of bubble equations that need to be evaluated increases linearly with $n$. Second, a higher number of centers means a higher number of degrees of freedom of the evolutionary algorithm, thus the population size and the number of offspring per generation should be increased accordingly, to optimize the algorithm's performance.

When run on a three-center configuration, the algorithm typically found of order 10 approximate solutions with fitness $\gtrsim10^{6}$ in around 10 hours (all run-times refer to a high-specification mainstream desktop machine). 
For a five-center configuration, typically around 13 hours runtime produced two solutions with fitness $\gtrsim10^{6}$. 
For a seven-center configuration, in 40 hours runtime the algorithm produced two solutions with fitness $\gtrsim 10^5$, including the one reported above. Each of these runs was performed initially with a cutoff of 10000 generations. Subsequently, longer runs were performed on the two examples presented earlier in this section to obtain the machine-precision solutions.

In light of the No Free Lunch Theorem, we have compared our algorithm with a random search on several examples, and observed it is always far superior, as follows. The random search is obtained by generating  $\texttt{offspring\_per\_generation}$ new individuals via  Eq.~\eqref{new individual} for $\texttt{generations}$ generations, and evaluating their fitness. We did so for different values of the standard deviation $\texttt{var\_pos}$. For large values of $\texttt{var\_pos}$, the relevant sampling space is too big, and the probability of finding good solutions is low. As we decrease $\texttt{var\_pos}$, the performance of the random search increases until an optimized value. Decreasing $\texttt{var\_pos}$ further results in a loss of performance, as the individuals are too close to the seed solution $\bar r_A$, and thus their fitness is of the same order of $f\big(\bar r_A\big)$. 

\begin{figure}[h!]
    \centering
    \includegraphics[width=10.5cm]{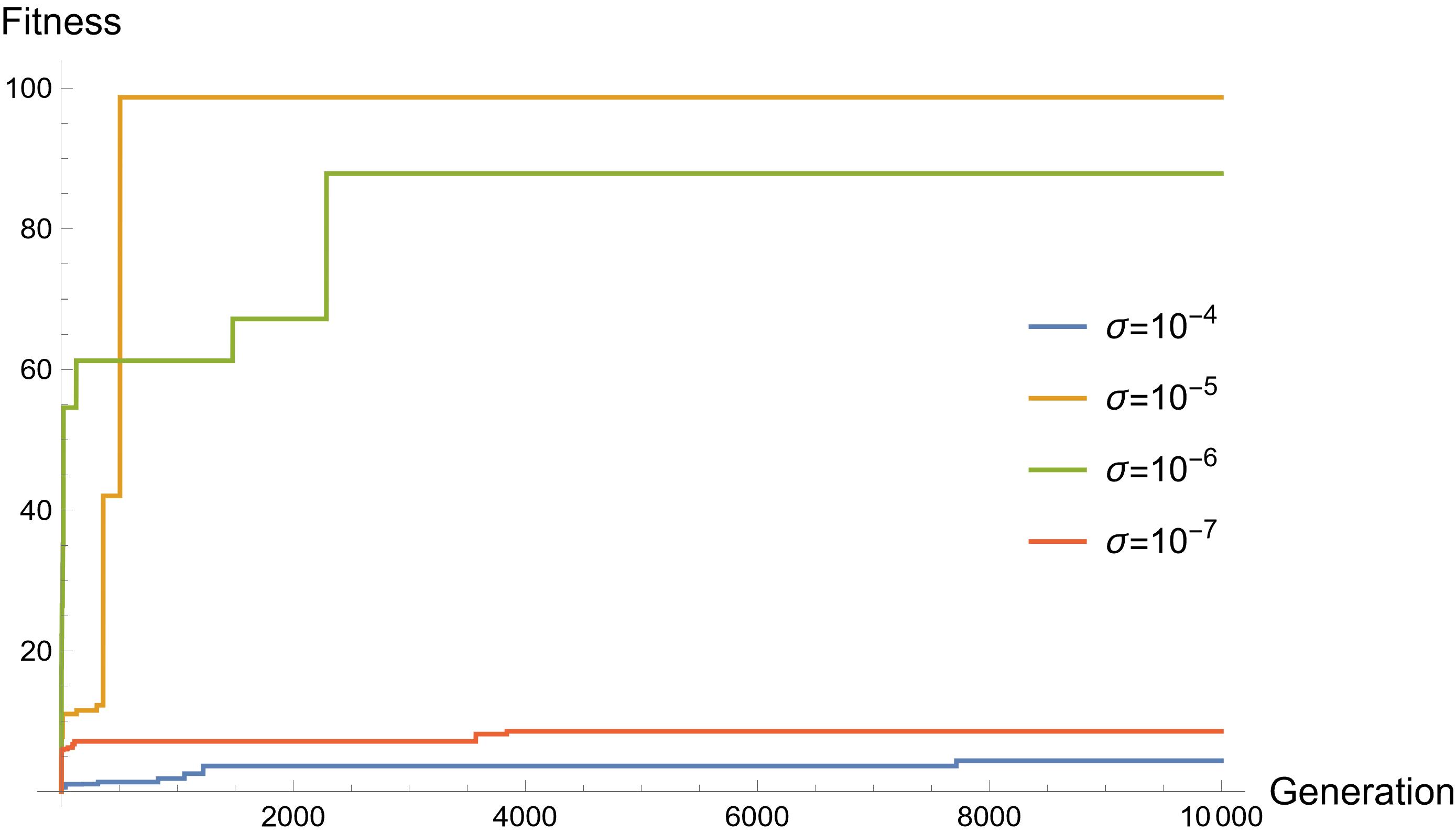}
    \caption{Random search over the initial configuration described in Tables~\ref{tab: 5 centers} and~\ref{tab: 5 centers fluxes}. We repeat the analysis for four different values of $\texttt{var\_pos}$, which are denoted with $\sigma$ in the plot's legend.}
    \label{fig:5 random}
\end{figure}
In all the examples we analysed, the random search provided an approximate solution with fitness no higher than around $10^2$.
For completeness, in Figure~\ref{fig:5 random} we present an example of such a random search for the five-center configuration discussed in~\ref{sec: 5 centers example}, with the values of $\texttt{generations}$ and $\texttt{offspring\_per\_generation}$ given in Table~\ref{tab: Hyperparameter 5 cent}.
This contrasts with the far superior performance of the evolutionary algorithm.

\section{Discussion}\label{sec: conclusions}

In this work we have presented an optimization algorithm, combining Bayesian optimization and an evolutionary algorithm, using which we have constructed numerical multi-center solutions with several centers arranged in generic configurations, satisfying all flux quantization constraints.

The Bayesian optimization algorithm generates a configuration in the scaling regime with appropriately large D1 and D5 charges, by optimizing over a subset of the flux parameters. The output of this first step is a starting configuration that is an approximate solution to the bubble equations~\eqref{bubble equation original}. The starting configuration contains two approximations: first, it was derived by solving the homogeneous form of the bubble equations; second, the flux parameters were rounded in order to respect the conditions of flux quantization.

The evolutionary algorithm uses the configuration generated in the first step as a starting point to generate approximate solutions to the full inhomogeneous bubble equations, with increasing precision. It does so by optimizing over the positions of the centers. By working with the positions of the centers, the distances between centers are always well-defined. The fact that we work with quantized fluxes and well-defined distances is a distinguishing advantage of our method over previous approaches.

In Section~\ref{results} we exhibited two examples of novel machine-precision numerical solutions, one with five centers and one with seven centers. Although we have used the algorithm primarily to construct configurations with three, five and seven centers, it can in principle be used to construct configurations with any number of centers in generic configurations, upon tuning the hyperparameters appropriately.

While the evolutionary algorithm significantly improves upon previous methods, naturally it does not always find machine-precision solutions, depending on the starting configuration. This can be for one of two reasons. First, it is a general limitation of evolutionary algorithms that they do not always find optimal solutions, i.e.~global minima of the fitness function: in the evolution process, the population might get stuck in a local minimum. It is well known that this problem is more serious when the number of degrees of freedom of the problem is low, as  the probability of having local minima decreases as the number of dimensions increase.

Second, given a set of poles in the harmonic functions describing the multi-center solution, it is not guaranteed that there exists a nearby exact solution to the bubble equations. For instance, by rounding the flux parameters, one could access a region of parameter space that does not admit solutions. 
It is thus possible that the evolutionary algorithm fails to find a sufficiently good solution simply because such a solution does not exist. 
Indeed, we found three-center configurations in which the evolutionary algorithm could not improve the fitness above a certain value, and a detailed analysis of one such example showed that a nearby exact solution did not exist. Despite these inherent limitations, most of the time the algorithm is successful at finding solutions.

As the number of centers grows, naturally more computational resources are required. It is worth noting that, in particular, the task of finding a good starting configuration requires significantly more computational resources.  This is because it can take several iterations for the Bayesian optimization algorithm to find scaling configurations with appropriately large supergravity charges $Q_1,Q_5$, together with a  positive-definite matrix $\mathcal{M}$ for absence of CTCs. So as the number of centers increases, a smaller fraction of potential starting configurations get passed on to the evolutionary algorithm.
In particular, apart from choosing the flux parameters that we initialize to have the same sign, as suggested in~\cite[Footnote~17]{Avila:2017pwi}, finding a good $\mathcal{M}$ is otherwise done by a random search. 
The probability of randomly selecting initial parameters that lead to an $\mathcal{M}$ with all positive eigenvalues decreases as the number of centers (and thus the dimension of $\mathcal{M}$) increases. Therefore, it would be interesting to explore more efficient ways of selecting parameters that lead to a positive-definite $\mathcal{M}$; we leave this task to future work.

To conclude, we have implemented a novel application of evolutionary algorithms and Bayesian optimization to the study of multi-centered solutions to supergravity, and have presented two state-of-the-art machine-precision numerical solutions. 
Our algorithm could be used to create larger families of numerical multi-center solutions, which could in turn be used in phenomenological models~\cite{Bena:2020see,Bianchi:2020bxa,Bah:2021jno}. 
The prospects for harnessing the power of computer science algorithms to solve physically interesting problems in String Theory and related fields appear bright, with an exciting future ahead.

\vspace{4mm}
\section*{Acknowledgements}

We thank Iosif Bena, Johan Bl{\aa}b{\"a}ck, \'{O}scar Dias, Bogdan Ganchev, Pierre Heidmann, Anthony Houppe, Daniel Mayerson, Rodolfo Russo, Kostas Skenderis, Marika Taylor, Nicholas Warner and Ben Withers for fruitful discussions.
The work of SR was supported by a Royal Society URF Enhancement Award.
The work of DT was supported by a Royal Society Tata University Research Fellowship.
We thank the IPhT, CEA Saclay and the University of Genoa for hospitality during the course of this work.


\vspace{0.5cm}


\bibliographystyle{utphys}      
\bibliography{microstates}       



\end{document}